# VORTEX SHEDDING DYNAMICS IN THE LAMINAR WAKE OF CONES

Michel Provansal[1] and Peter A. Monkewitz[1,2]
[1] IRPHE Aix-Marseille University, 49 rue F.Joliot-Curie, B.P. 146
13384 Marseille cedex 13 FRANCE
[2] Laboratory of Fluid Mechanics, EPFL, CH-1015 Lausanne SWITZERLAND

**Keywords**: *wakes, cones and cylinders, image processing, space-time analysis*

ABSTRACT

*Experiments on two cones of different taper ratios have been performed in the periodic Reynolds number range between 40 and 180. The visualizations of the plan view of the wake with hydrogen bubbles allow to determine local instantaneous frequencies, wavelengths and shedding angles from digital movie. The shedding frequency adjusts in a stepwise manner to the continuous variation of the cone diameter.*

## 1 INTRODUCTION

The purpose of this work is to analyze the vortex shedding dynamics in the near wake of cones. This configuration has been previously studied by different authors [1,2,3,4] . The main feature of the flow is the existence of discrete cells along the spanwise direction where the vortex shedding frequency is constant. The characteristic angle of vortex shedding lines has been determined by Radon transform of different images. However some problems are unclear. Is it possible to control the vortex dynamics from ends as in the case of the cylinder? Do the boundaries between cells move with time? Our experiments and image processing address these questions and bring new information.

## 2 EXPERIMENTAL SET-UP

### 2.1 Water channel and cones

Experiments have been performed in a free surface water channel. The dimensions of the water channel are 2 m in the streamwise direction , 0.6 m along the vertical one and 0.5 m in the transverse horizontal axis. The upstream velocity of the flow might be varied from 0.05 m/s to 0.5 m/s with uniform velocity profile and low turbulence level. The geometry is the following Oxyz with axis Ox along the horizontal streamwise direction, Oz vertical up and Oy horizontal transverse direction. The origin O has been taken in the middle of the cone or of the cylinder. Two cones made of polished brass with taper ratios $3.2 \ 10^{-3}$ and $6.7 \ 10^{-3}$ have been used. Along the span of the small cone (L = 0.29 m) the diameter changes from d = 1.1 mm to D = 3,05 mm with an average value $d_{av}$ = 2.075 mm. The great cone length is L' = 0,59 m ; its diameter varies from d = 1,7 mm to D = 3,6 mm. The cones are held by two metallic bar linked to a frame which can be rotated in the vertical plane xOz. At the ends of the cones , two plates made of "plexiglass" have a thin thickness (5 mm) and a length 5 cm (cf. Fig. 1). In almost all the experiments, the upstream flow velocity was around 3 cm/s. It was selected such that the Reynolds number is in the periodic vortex shedding regime and less than 200 where the three-dimensional modes A or B [5,6] can appear in the circular cylinder wakes. The control of end effects was monitored by a small water stream created by suction through two pipes (visible down and up of Fig. 1) . Special care has been paid to reduce the mean spanwise velocity as it was done previously for the circular cylinder configuration [7] .

## 2.2 Visualization method

A small metallic wire has been located 10 cm upstream of the cone to produce a vertical line of small bubbles. The diameter of the wire (50 µm) was chosen to get an uniform visualization of the flow. An electric voltage of 30 to 40 volts has been applied between the wire and a metallic plate set 40 cm downstream. Without any cone, the stability of the flow at low velocity has been checked from observation of the bubble plane xOz. The relative transverse position of the wire and of the cone (or cylinder) has been modified to feed a single vortex side of the wake along the span (Fig. 2) (method already used in air [8]). A blue laser sheet ($\lambda = 0,488$ µm, power 2 watts) or the white light of a halogen spot (300 watts) illuminates the wake downstream. The laser ray is focused by an optic fiber downstream of the bluff body [9]. A cylindrical lens creates a slowly diverging sheet. Visualizations have been recorded with a numerical monitor "Camescop JVC" during a time of 10 mn which corresponds to thousands of vortex shedding for the cone.

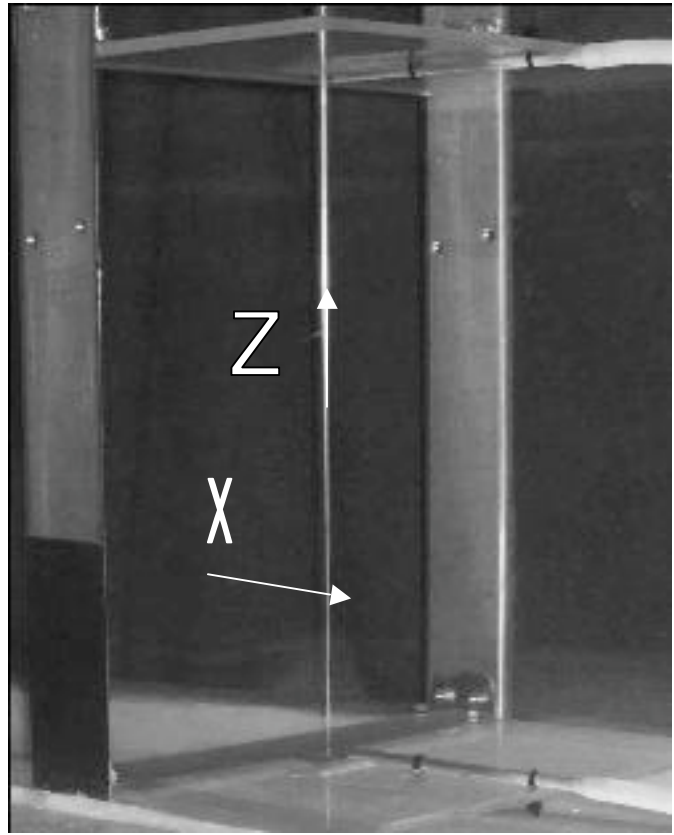

Fig. 1  View of the longer cone in the water channel

The pictures sequenced at the video frequency (25 Hz) have been processed to reduce the noise and artifacts. All the images have been stored with the standard RGB. The illumination by the blue laser light is such that some noisy bubbles diffuse white light. The Red filter is sensitive to this wavelength band and might be used to suppress some parasite blue reflections. After different tests, we have decided to use the grayscale version of the image RGB which includes the maximum of details. Different filters have been used to suppress noisy bubbles and to increase the signal to noise ratio. Another operation has been used to avoid excess of luminosity of pixel on their neighbors and to get a better contour of images. In the final version, images are set black and white with a discrimination of the lower noisy intensity.

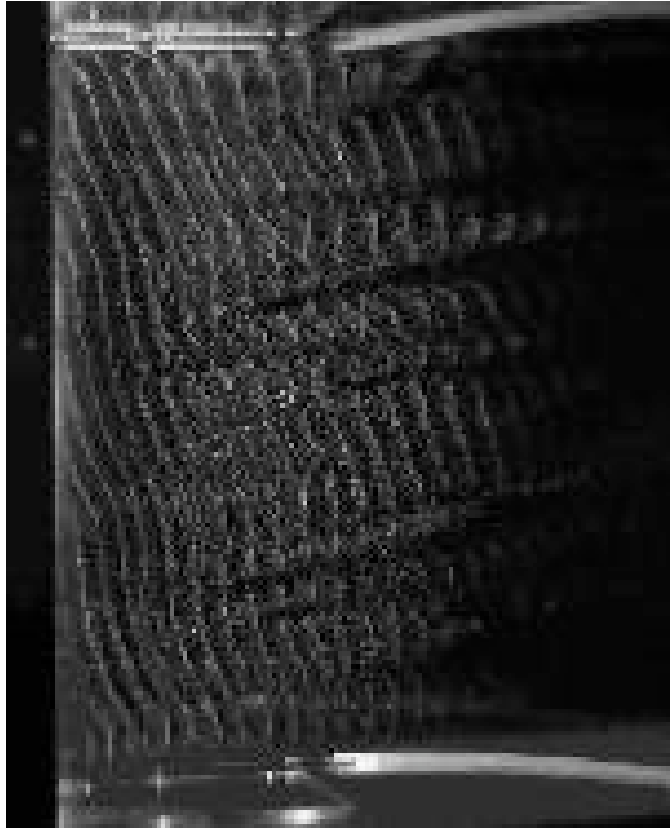

Fig. 2 Visualizations of the vortex lines in the wake of the longer cone

## 3 ANALYSIS OF THE VORTEX SHEDDING ANGLES BY RADON TRANSFORM

### 3.1 The Radon Transform

The analysis of characteristic vortex angles has be done by the Radon Transform [10]. In this operation, an image is swept by a beam for all the incidence angles and the transform gives the result of the integration of all the material along each ray of the beam. This Transform involves two parameters : the angle of incidence $\theta$ perpendicular to the irradiating beam and the distance Xr which separates this beam from the center (cf. Fig. 3).

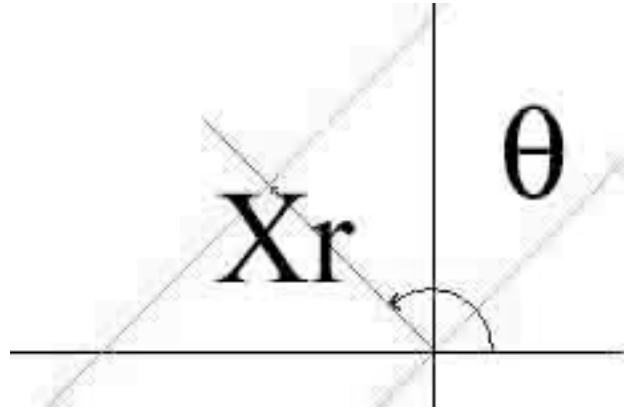

Fig. 3 Definition of the parameters of the Radon Transform

This tool is widely used in medicine . For instance, the inverse Radon Transform allows to reconstruct the image of a body irradiated by different parallel X rays.

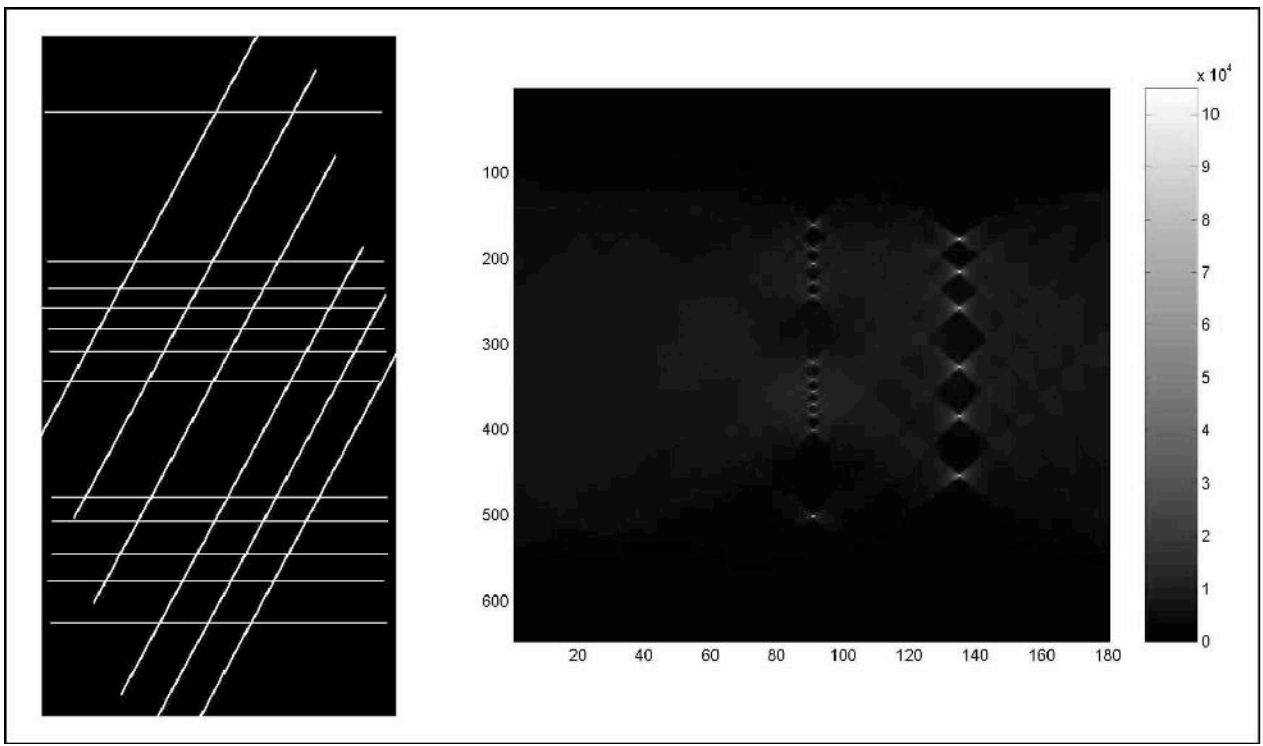

Fig. 4 Radon Transform of a test picture (right) and the results of the transform (left with angles in degrees along the horizontal axis and the distance along the vertical axis).

We have tested the validity of our algorithm developed under Matlab on different images (Fig. 4). We have drawn a single horizontal line, a group of six parallel horizontal lines, another one of five horizontal lines and a set of six oblique lines The horizontal lines of the picture correspond to the normal angle θ = 90 ° of the Radon Transform on the left . The oblique lines give an angle near 135° with a pretty good accuracy.

## 3.2 Field of the vortex shedding angles θ

The evolution of the angle θ( z,t) with space and time has been investigated. The analysis has been performed on a strip parallel to the axis of the cone in the near wake where the vortex shedding lines are well developed without any diffusion of bubbles. The strip is composed of blocks with a strong overlapping (50 %) to assess a compromise between the increase of the number of blocks and a drastic reduction of the size.

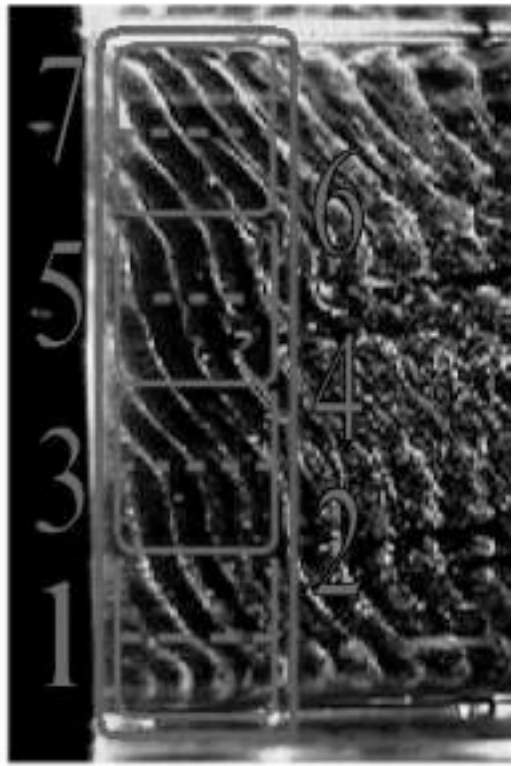

Fig.5 Definition of a strip and of different blocks

On Figure 5, the dimensions of the blocks and of the strip have been oversized to make it visible. In fact, the width of the strip is close to the vortex shedding wavelength around five diameters. The height of each block is hundredth of the strip height. For every block, at each time step, the Radon transform has an angle maximum and this value has been associated to the space and time set. The three-dimensional view of the angle (Figure 6) shows clearly the locations and the time where dislocations between vortex lines occur. It is obvious that the boundary between two cells move with time and space. The research of long time behavior requires a too large and complex computation to be made by Radon transform .

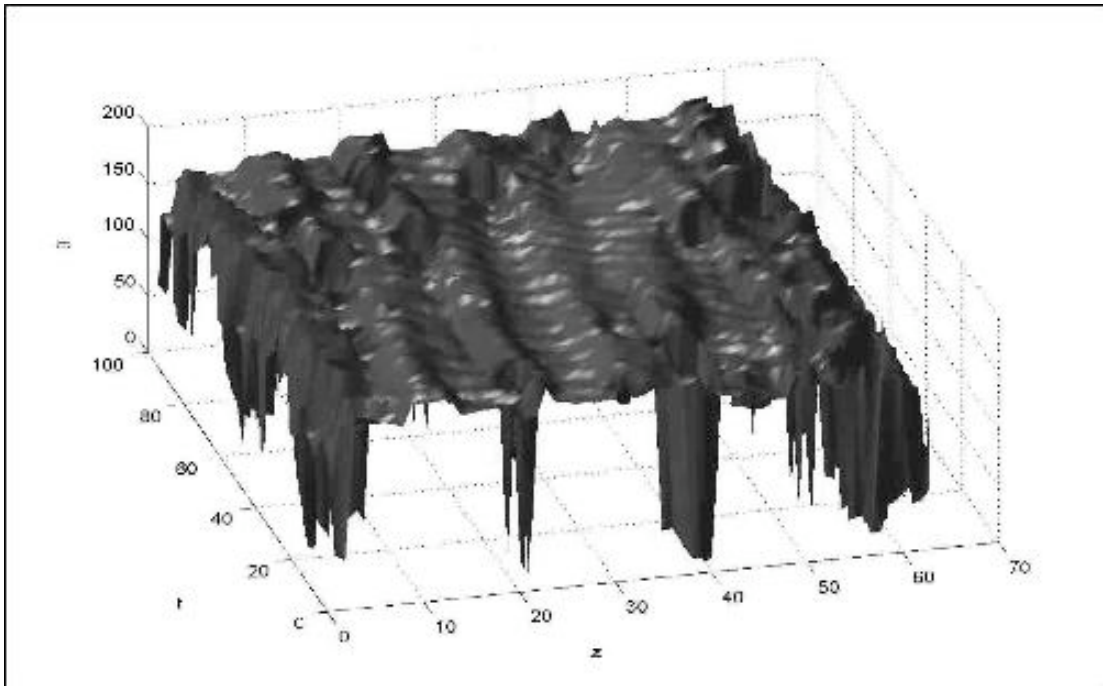

Fig. 6 Analysis of the vortex shedding angle θ(z, t) in degrees as function of space of z (pixels) and t time (number/25 s).

## 4 RESULTS AND DISCUSSION

### 4.1 Determination of temporal periods T(z,t)

The time period T(z,t) needs to be determined at each time for every location along the axis of the cone. The observation of the time period field reveals singularities similar to the orientation θ(z,t).

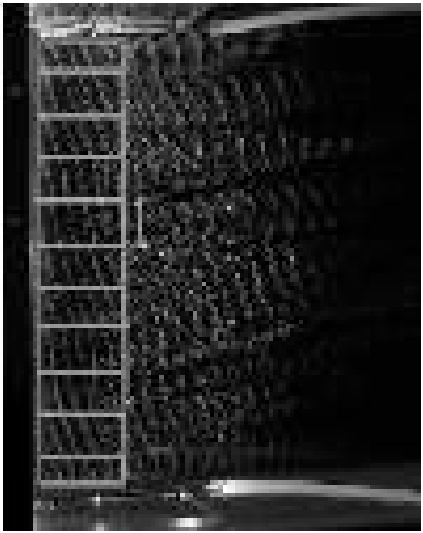
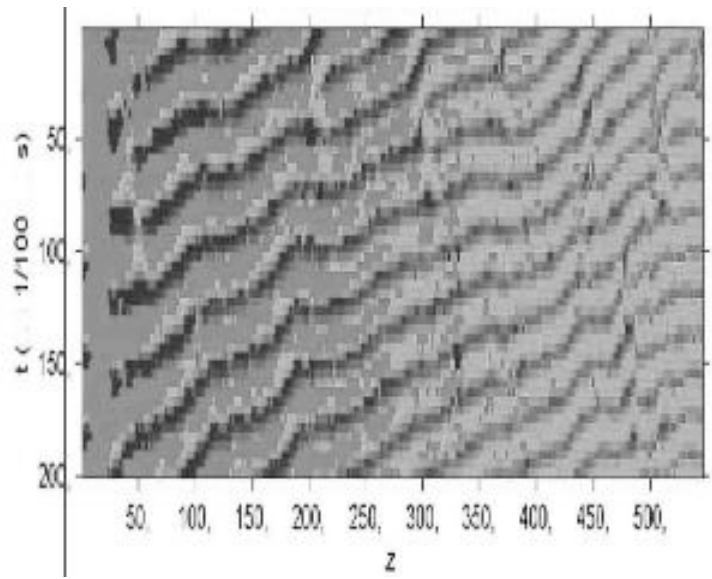

Fig. 7 Decomposition of strings .          Fig.8 T(z,t) for the long cone (thin end of the cone on the right)

We have developed an algorithm which gives the period from auto correlation of strings along the time (Fig. 7) and processed on each image. The main difficulty is here to get the knowledge of the period for every point (Fig. 8) to take into account the dynamics of the vortex.

**4.2 Space –time reconstruction of the wake and spectral analysis**

A line parallel to the axis of the cone has been selected in the near wake where the contrast is good at any time. Figure 9 exhibits the evolution of successive intensities of this line along the span.

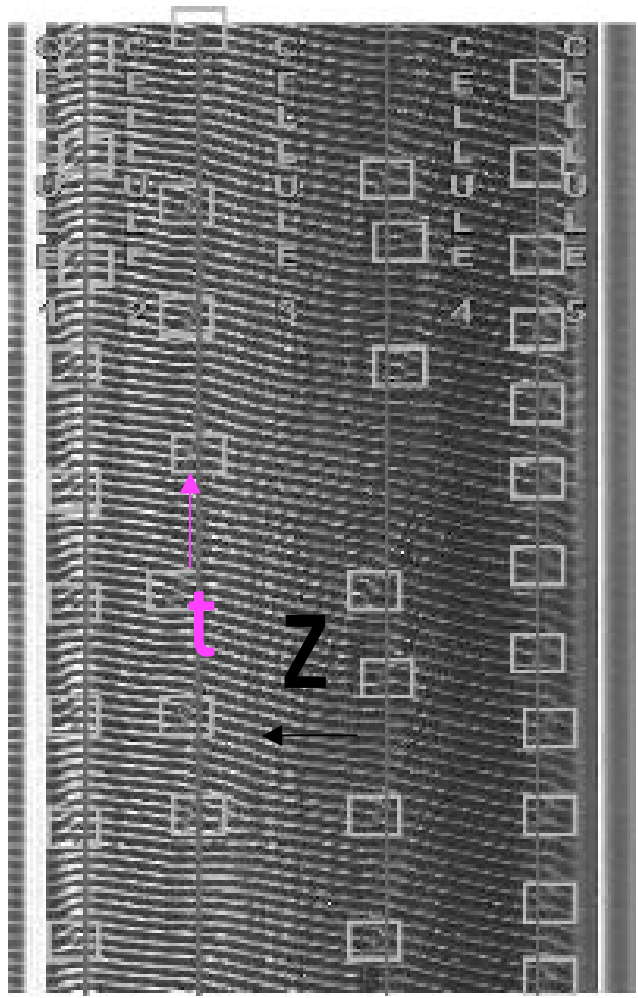

Fig. 9 Space –time diagram of the near wake (thin end of the cone on the right, $Re_{average}$ = 92 suction on the ends).

Five cells coexist as in the experimental work of Noack [11] or in numerical simulations of cone wakes [12,13]. The dislocations have been pointed out by clear boxes . Their distribution is more regular in space and in time near the ends of the cone or near the end plates. The black or straight lines of the Figures 2 and 9 lead to the hypothesis that there is some propagation towards the thin end of the cone. The evolution of the dislocations has been more precisely described in [14,15].

The results obtained for the vortex shedding angle θ(z,t) and for the period T(z,t) show that a complete investigation must take into account the dynamics of the dislocations with time and space. The spectral analysis of the fluctuation of intensity with vortex lines has been performed for each spanwise location of the cone (Fig. 10). The discrete values of the frequency, organized as the stairs of Gaster [2] are typical of different cells. We underline the fact that such a long time averaging is insufficient and might lead to misunderstandings.

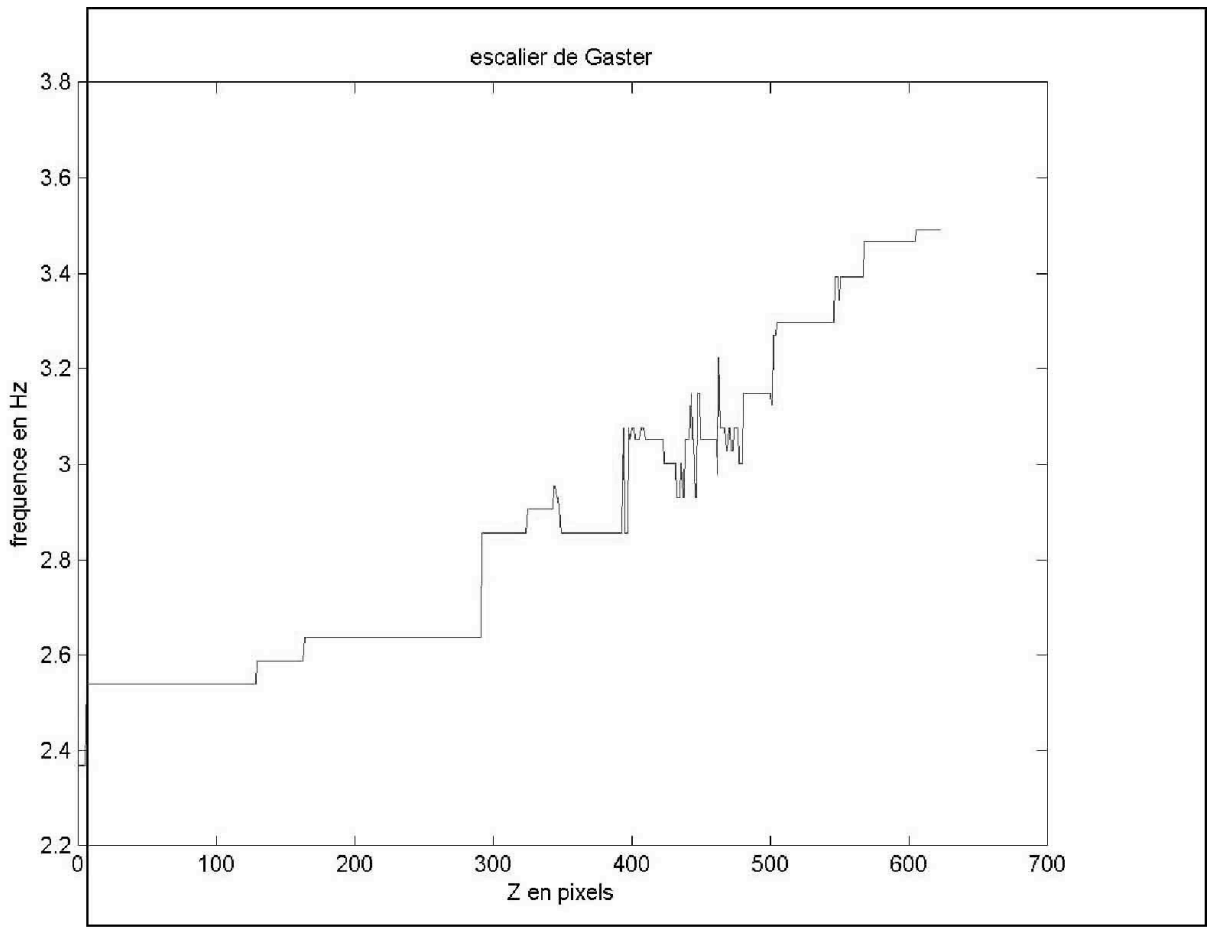

Fig. 10 Discrete variation of the frequency along the span of the cone( thin end on the right).

## 5 CONCLUSION AND OUTLOOK

The wake of a cone has been visualized and the images have been processed by different methods. The vortex shedding has been characterized by angles deduced from a Radon transform and by a space–time analysis. The time period of oscillation has been measured on every point of the cone by correlation. The Fast Fourier Transform has been used to get the average frequency of vortex shedding process. A stair case variation has been observed. However, such a description doesn't explain all the phenomena and a careful investigation of the frequency is necessary.

We are grateful to Thomas Leweke who has made available the water channel and to Jacky Minelli and all the team of mechanics laboratory of I.R.P.H.E. for their technical assistance. P.Monkewitz was supported by grants from CNRS and City of Marseills while on sabbatic leave.